\documentclass[conference]{IEEEtran}
\IEEEoverridecommandlockouts
\usepackage{amsmath,amssymb,amsfonts}
\usepackage{amsthm}
\usepackage[hyphens]{url}
\usepackage{hyperref}
\usepackage[most]{tcolorbox}
\usepackage{algorithm}
\usepackage{algpseudocode}
\usepackage{graphicx}
\usepackage{textcomp}
\usepackage{siunitx}
\usepackage{bm}

\usepackage{multirow}
\usepackage{multicol}
\usepackage{xcolor}
\usepackage{enumitem}
\usepackage{csquotes}
\usepackage{tcolorbox}
\usepackage{hyperref}
\usepackage{booktabs}
\usepackage{tikz}
\usetikzlibrary{positioning}
\usepackage{natbib}
\newcommand{\EpsSet}{\{0.1,\,0.5,\,1,\,2,\,5\}}
\usepackage[T1]{fontenc}
\usepackage{newtxtext,newtxmath}

\begin{document}

\title{Federated Survival Analysis with Node-Level Differential Privacy: Private Kaplan-Meier Curves
}

\author{\IEEEauthorblockN{1\textsuperscript{st} Narasimha Raghavan Veeraragavan}
\IEEEauthorblockA{\textit{Cancer Registry of Norway} \\
\textit{Norwegian Institute of Public Health}\\
Oslo, Norway \\
Narasimha.Raghavan.Veeraragavan@fhi.no}
\and
\IEEEauthorblockN{2\textsuperscript{nd} Jan~F.~Nygård}
\IEEEauthorblockA{\textit{Cancer Registry of Norway}\\
\textit{Norwegian Institute of Public Health}\\ 
\textit{Oslo, Norway} \\
\textit{and The Arctic University of Norway}\\
Tromsø, Norway \\
Jan.Franz.Nygard@fhi.no}
}

\maketitle
\renewcommand{\thefootnote}{\fnsymbol{footnote}}
\footnotetext[1]{© 2025 IEEE. Personal use of this material is permitted. 
Permission from IEEE must be obtained for all other uses, in any current or 
future media, including reprinting/republishing this material for advertising 
or promotional purposes, creating new collective works, for resale or 
redistribution to servers or lists, or reuse of any copyrighted component 
of this work in other works. This is the author’s accepted version of the 
paper. The final version of record will appear in \textit{Proceedings of the 
IEEE International Conference on Federated Learning Technologies and 
Applications (FLTA 2025)} and will be available at IEEE Xplore.}
\renewcommand{\thefootnote}{\arabic{footnote}}
\begin{abstract}
We investigate how to calculate Kaplan–Meier survival curves across
multiple health-care jurisdictions while protecting patient privacy with
node-level differential privacy.
Each site discloses its curve only once, adding Laplace noise whose
scale is determined by the length of the common time grid; the server
then averages the noisy curves, so the overall privacy budget remains
unchanged.
We benchmark four one-shot smoothing techniques: Discrete Cosine
Transform, Haar Wavelet shrinkage, adaptive Total-Variation
denoising, and a parametric Weibull fit on the NCCTG lung-cancer
cohort under five privacy levels and three partition scenarios
(uniform, moderately skewed, highly imbalanced).
Total-Variation gives the best mean accuracy, whereas the frequency-
domain smoothers offer stronger worst-case robustness and the Weibull
model shows the most stable behaviour at the strictest privacy
setting.
Across all methods the released curves keep the empirical log-rank
type-I error below fifteen per cent for privacy budgets of
0.5 and higher, demonstrating that clinically useful survival
information can be shared without iterative training or heavy
cryptography.
\end{abstract}

% Add IEEE copyright disclaimer as a footnote on the first page

%=====================================================================
\section{Introduction}\label{sec:intro}
%=====================================================================
Time-to-event outcomes such as \emph{overall survival}, \emph{progression free
survival}, or \emph{time to hospital readmission} are central to
clinical trials and epidemiological studies.
The \emph{Kaplan–Meier} (KM) estimator~\cite{kaplan} is the
work-horse in this domain: a non-parametric, step-function estimate of
the survivor function \(S(t)=\Pr(T>t)\) that supports direct visual
inspection and classical inference tools such as the log-rank test.
Because many diseases are rare or treated in specialised centres,
reliable KM curves often require pooling data across multiple
health-care institutions and jurisdictions.

Publishing even an \textit{aggregate} KM curve poses privacy risks:
reconstruction attacks can infer individual events from small step
heights~\cite{rogula2022method,guyot2012enhanced,wei2017reconstructing}.
Consequently, regulations such as the GDPR~\cite{GDPR} restrict
cross-site data sharing.

Three strands of work tackle this problem.
(i) \textit{Secure computation} protocols
(homomorphic encryption, garbled circuits) merge event counts without
decryption~\cite{veeraragavan2024mph,froelicher2021};
however, the final curve is released in the clear and the cryptographic
overhead is high.
(ii) \textit{Centralised differential privacy (DP)} adds Laplace noise
to statistics held by a trusted curator.
Gondara and Wang~\cite{gondara2020differentially} perturb the at-risk
and event counts at each distinct time and rebuild the curve; this approach is referred 
as \textit{DP-Matrix}.  
Rahimian \textit{et al.}~\cite{rahimian2024private} introduce two
variants for equi-spaced grids: \textit{DP-Surv}, which adds Laplace
noise only to the first few discrete-cosine-transform coefficients,
and \textit{DP-Prob}, which perturbs the discrete hazard directly and
renormalises it.  
A more recent method combines a time-indexed noise schedule with
dynamic clipping and rolling-window smoothing
\cite{raghavan2024differentially}.  
All of these techniques assume the raw data remain in a single
repository.
(iii) The only \textit{federated, node-level DP} solution to date is
\textsc{Collaborative DP-KM}~\cite{rahimian2024private}, which extends
DP-Surv and DP-Prob to multiple sites but evaluates a single smoother, under one
privacy budget.

We introduce an entirely \emph{one-shot}, node-level DP pipeline:
\begin{enumerate}[leftmargin=1.8em,label=(\roman*)]
  \item Each health-care institution evaluates its KM vector on a public time grid
        $\tau$.
  \item A single Laplace draw (scale $1/|\tau|$) is processed by one of
        four smoothers: Discrete Cosine Transform (\textsc{DCT})~\cite{dct},
        Haar \textsc{Wavelet}~\cite{wavlet}, adaptive Total-Variation (\textsc{TV})~\cite{tv},
        or a parametric \textsc{Weibull} fit~\cite{weibull}.
  \item The coordinator averages the noisy curves; 
\end{enumerate}
The last three smoothers are, to our knowledge, new in the DP
literature for KM curves, and no previous work has compared their
utility, robustness to data skew, and statistical fidelity.

Focusing on the NCCTG lung-cancer cohort (\(n=228\))~\cite{R_survival_lung} we sweep five
privacy budgets \(\varepsilon\in\{0.1,0.5,1,2,5\}\) and three
partitioning schemes (uniform, moderate skew (60–20–20), highly imbalanced skew (90–5–5)).
We assess (i)~\emph{utility vs.\ privacy} via mean absolute error,
(ii)~\emph{robustness to data skew}, (iii)~\emph{method ranking} by
average ordinal score, and (iv)~\emph{statistical fidelity} via the
log-rank test, repeating every configuration 100 times with
independent noise seeds.
\begin{tcolorbox}[colback=blue!5!white,colframe=white,breakable]
The summary of our contributions are as follows: 
\begin{itemize}[leftmargin=1.7em]
  \item We introduce three \emph{novel} one-shot DP smoothers
        (Haar-\textsc{Wavelet}, adaptive \textsc{TV}, parametric
        \textsc{Weibull}) for KM curves and cast \textsc{DCT} in a
        federated node-level DP setting.
  \item We propose an adaptive grid rule that balances time resolution
        against privacy noise and enforce legality via a monotone
        projection.
  \item We deliver the first head-to-head evaluation of DP-KM
        smoothers under varying privacy budgets and data-imbalance
        scenarios, showing that useful survival information can be
        shared with node-level \((\varepsilon,0)\)-DP at
        \(\varepsilon\!\ge\!0.5\).
  \item All code and plotting scripts are released for reproducibility
        and future extensions: \url{https://github.com/CancerRegistryOfNorway/DifferentiallyPrivateKaplanMeier.git}
\end{itemize}
\end{tcolorbox}
%=====================================================================
\section{Methodology}\label{sec:method}
%=====================================================================

This section (i) fixes notation, (ii) outlines the federated DP–KM
pipeline, and (iii) details four node-level smoothing mechanisms
(Algorithms \ref{alg:dct}–\ref{alg:weibull}).

%---------------------------------------------------------------------
\subsection{Notation}\label{sec:notation}
%---------------------------------------------------------------------
Table \ref{tab:global-notation} gathers the \emph{global} symbols
shared across Secs.~\ref{sec:pipeline}–\ref{sec:algs}.
Algorithm-specific symbols (e.g.\ wavelet coefficients $w_\ell$,
TV weight $\lambda$) are listed later in
Table \ref{tab:algo-notation}.

\begin{table*}[t]
\centering\small
\caption{\textbf{Global notation} used throughout
         Secs.~\ref{sec:notation}–\ref{sec:algs}.
         Per-algorithm symbols appear in Table~\ref{tab:algo-notation}.}
\label{tab:global-notation}
\begin{tabular}{@{}ll@{}}
\toprule
\textbf{Symbol} & \textbf{Meaning} \\
\midrule
$M$                            & Number of participating nodes \\[2pt]
$D_i=\{(t_{i,r},\delta_{i,r})\}_{r=1}^{n_i}$ & Event/censor times at node $i$ \\[2pt]
$n_i$                          & Local sample size of node $i$ \\[2pt]
$\tau=\{t_1<\dots<t_K\}$       & Common evaluation grid (Section~\ref{sec:grid-choice}) \\[2pt]
$K$                            & Grid length, set by Eq.~\eqref{eq:grid} \\[2pt]
$\widehat S_i$                 & Node-level Kaplan–Meier vector on $\tau$ \\[2pt]
$\widehat S^{\text{fed}}$      & Federated DP-KM curve after aggregation \\[2pt]
$\varepsilon$                  & \emph{Global} privacy budget (node $i$ uses $\varepsilon_i=\varepsilon/M$) \\[2pt]
$\Delta$                       & $\ell_\infty$ sensitivity, $\Delta=1/K$ (Sec.~\ref{sec:sens}) \\[2pt]
$b=\Delta/\varepsilon_i$       & Laplace scale for node $i$ \\[2pt]
$R$                            & Number of Monte-Carlo repetitions per setting \\
\bottomrule
\end{tabular}
\end{table*}

%---------------------------------------------------------------------
\subsection{Pipeline overview}\label{sec:pipeline}
%---------------------------------------------------------------------
For every experimental configuration
$\langle\text{partition},m,\varepsilon\rangle$:

\begin{enumerate}[leftmargin=2em,label=(\arabic*)]
  \item \textbf{Local KM computation.}  Node $i$ evaluates its raw KM
        vector $\widehat S_i\in\mathbb{R}^K$ on $\tau$.
  \item \textbf{DP smoothing.}  One mechanism
        $m\in\{\textsc{DCT},\textsc{Wavelet},\textsc{TV},\textsc{Weibull}\}$
        is applied with Laplace scale $b=\Delta/\varepsilon_i$
        (Algorithms \ref{alg:dct}–\ref{alg:weibull}).
  \item \textbf{Post-processing.}  The noisy output is clipped to
        $[0,1]$ and made monotone by the cumulative minimum; post-processing
        costs no additional privacy.
  \item \textbf{Secure aggregation.}  The coordinator publishes
        $\widehat S^{\text{fed}}
         =\frac1M\sum_{i=1}^{M}\widehat S_i^{\text{DP}}$,
        inheriting node-level $(\varepsilon,0)$-DP.
\end{enumerate}

%---------------------------------------------------------------------
\subsection{Sensitivity and noise calibration}\label{sec:sens}
%---------------------------------------------------------------------
Deleting or adding a single patient changes at most one KM step by
$1/K$, so the node-level $\ell_\infty$ sensitivity is
$\Delta = 1/K$.
All Laplace perturbations therefore use
$b=\Delta/\varepsilon_i$.

%---------------------------------------------------------------------
\subsection{Adaptive evaluation grid}\label{sec:grid-choice}
%---------------------------------------------------------------------
With total patient count $n=\sum_{i=1}^{M}n_i$, the grid length $K$ is
\begin{equation}
  K \;=\; \min\!\Bigl\{\,\bigl\lceil\rho\,n\bigr\rceil,\;K_{\max}\Bigr\},
  \qquad 0<\rho\le 1,\;K_{\max}\in\mathbb{N},
  \label{eq:grid}
\end{equation}
where $\rho$ is a \emph{grid-density factor} and $K_{\max}$ a safety
cap.  Choosing $K$ directly tunes the Laplace scale
$b=\Delta/\varepsilon_i$.

\subsubsection{Why \texorpdfstring{$\rho$}{rho} matters}\label{sec:grid-tuning}
A coarse grid (small $\rho$) yields fewer evaluation points and thus a
larger $\Delta$, injecting more noise at each point but at fewer
locations.  A dense grid (large $\rho$) does the opposite.
An optimal $\rho^\star$ balances temporal resolution against total
noise energy.  In practice we run a lightweight privacy-free pilot
grid search (Sec.~\ref{sec:grid-tuning}) to select $\rho$ and
$K_{\max}$ before the main study.

%---------------------------------------------------------------------
\subsection{Post-processing: cumulative minimum}\label{sec:postproc}
%---------------------------------------------------------------------
After noise injection each smoother applies  
\[
  \widetilde S \;\leftarrow\;
  \operatorname{cummin}\!\bigl(\operatorname{clip}(\widetilde S,0,1)\bigr),
\]
where \texttt{cummin} replaces every entry by the minimum of all
preceding ones.  This deterministic projection enforces
\(1=S(t_1)\ge\dots\ge S(t_K)\ge 0\) without consuming privacy budget.

%---------------------------------------------------------------------
\subsection{DP smoothing mechanisms}\label{sec:algs}
%---------------------------------------------------------------------
Table \ref{tab:algo-notation} defines the symbols used exclusively in
Algorithms \ref{alg:dct}–\ref{alg:weibull}.  Global quantities have
already been introduced.

\begin{table*}[t]
\centering\small
\caption{Notation inside Algs.~\ref{alg:dct}–\ref{alg:weibull}.}
\label{tab:algo-notation}
\begin{tabular}{@{}ll@{}}
\toprule
Symbol & Meaning \\
\midrule
$c,\,c'$              & DCT coefficients before / after noise \\
$w_\ell$              & Haar-wavelet coefficient at level $\ell$ \\
$\text{DCT},\text{IDCT}$ & Forward / inverse discrete-cosine transform \\
$\text{HaarDecompose},\text{HaarReconstruct}$ & Wavelet analysis / synthesis \\
$\text{Laplace}(0,b)$ & I.i.d.\ noise with scale $b=\Delta/\varepsilon_i$ \\
$\lambda_{0}$         & Base TV regularisation weight (tuned on a pilot run) \\[0.25em]
$n_{0}$               & Pivot node size used in the scaling of $\lambda(n)$ \\[0.25em] 
$\lambda(n)$          & Adaptive TV weight \\ 
$\alpha$              & Size exponent in $lambda(n)$ \\
                     
$\|\,\cdot\,\|_{\mathrm{TV}}$ & One-dimensional total-variation seminorm \\
$x$                       & Candidate vector in the TV objective;one entry per grid point \(t_j\)\\[0.25em]
$\hat x$              & TV-denoised vector \\
$k,\,\lambda$         & Shape / scale of the Weibull model $\left(S(t)=\exp[-(t/\lambda)^{k}]\right)$ \\
                        \(S(t)=\exp[-(t/\lambda)^{k}]\) \\
\(\operatorname{clip}(x,0,1)\) & Entry-wise truncation to \([0,1]\) \\
\(\operatorname{cummin}(x)\)   &
  Cumulative minimum \([x_1,\min\{x_1,x_2\},\dots]\) \\
\bottomrule
\end{tabular}
\end{table*}

%----------------------------  DCT  ---------------------------------%
\begin{algorithm}[t]
\caption{\textsc{DCT} smoother (node $i$)}
\label{alg:dct}
\begin{algorithmic}[1]
\State $c \gets \text{DCT}(\widehat S_i)$
\State $c' \gets c + \text{Laplace}(0,\Delta/\varepsilon_i)$
\State $\widetilde S \gets \text{IDCT}(c')$
\State \Return clip + cummin
\end{algorithmic}
\end{algorithm}

\textbf{\textsc{DCT}.}  
Transforming to the cosine basis concentrates signal energy in the
first few coefficients; Laplace noise therefore attenuates high
frequencies more strongly, acting as an implicit low-pass filter.

%--------------------------- Wavelet --------------------------------%
\begin{algorithm}[t]
\caption{\textsc{Wavelet} smoother (node $i$)}
\label{alg:wavelet}
\begin{algorithmic}[1]
\State $\{w_\ell\}\gets\text{HaarDecompose}(\widehat S_i)$
\ForAll{$\ell$}\;$w_\ell\!\leftarrow\!
        w_\ell+\text{Laplace}(0,\Delta/\varepsilon_i)$\EndFor
\State $\widetilde S \gets\text{HaarReconstruct}(\{w_\ell\})$
\State \Return clip + cummin
\end{algorithmic}
\end{algorithm}

\textbf{\textsc{Wavelet}.}  
The Haar basis captures both global level and local drops; adding
independent Laplace noise to every coefficient preserves sharp early
events while damping late-time fluctuations.

%-----------------------------  TV  ---------------------------------%
\begin{algorithm}[t]
\caption{\textsc{TV} smoother (node $i$)}
\label{alg:tv}
\begin{algorithmic}[1]
\State $\lambda(n)\;=\;\lambda_{0}\!\left(\frac{n}{n_{0}}\right)^{\alpha}
                    \sqrt{\ln(n+1)}$

\State $\hat x\gets\arg\min_x\|x-\widehat S_i\|_2^{2}
                     +\lambda(n)\|x\|_{\rm TV}$\hfill{[Condat ’13]}
\State $\widetilde S\!\leftarrow\!
       \hat x+\text{Laplace}(0,\Delta/\varepsilon_i)
       -\text{mean}(\cdot)$
\State \Return clip + cummin
\end{algorithmic}
\end{algorithm}

\textbf{\textsc{TV}.}  
Total-variation denoising imposes a piece-wise constant prior that
retains the step-function nature of KM curves; the adaptive
$\lambda(n)$ prevents over-smoothing very small nodes.

%--------------------------- Weibull -------------------------------%
\begin{algorithm}[t]
\caption{\textsc{Weibull} smoother (node $i$)}
\label{alg:weibull}
\begin{algorithmic}[1]
\State Fit shape $k$ and scale $\lambda$ by log–log regression
\State $k\!\leftarrow\!k+\text{Laplace}(0,\Delta/\varepsilon_i)$
\State $\lambda\!\leftarrow\!\lambda+\text{Laplace}(0,\Delta/\varepsilon_i)$
\State $\widetilde S(t_j)=\exp[-(t_j/\lambda)^k]$
\State \Return clip + cummin
\end{algorithmic}
\end{algorithm}

\textbf{\textsc{Weibull}.}  
Perturbing the two parameters of a fitted Weibull model gives a
lightweight private surrogate; performance degrades when the true
hazard is markedly non-Weibull.

\medskip
After the common clip + cummin projection, every mechanism guarantees
one-shot $(\varepsilon_i,0)$ node-level DP for the federated
Kaplan–Meier estimator.

\section{Experiments}
In this section we: (a) describe the experimental setup, (b) pose the concrete research questions (RQs) that guide the study, (c) define the metrics that operationalise each RQ, and (d) describe the repetition protocol of the experiments. 

\subsection{Setup}\label{sec:setup}
\paragraph{Dataset.}
We use the publicly-available \textbf{NCCTG Lung‐Cancer} cohort, which
records overall survival following chemotherapy.  
The data consist of right-censored event times \(T\) (death or last
follow-up) and an event indicator
\(\delta\in\{0,1\}\) (\(1=\) event observed, \(0=\) censored).

\paragraph{Federated partitions}
To simulate a realistic multi-institutional setting, we partitioned the NCCTG lung cancer dataset into three sites under three distinct scenarios: (i) uniform split, where patients were evenly distributed across sites; (ii) moderately skewed split, where one site contained approximately half of the patients while the remaining sites shared the rest evenly; and (iii) highly imbalanced split, where a single site contained the majority of the patients and the others only small fractions. This design aims to capture typical heterogeneity observed in federated health-care consortia, where site sizes often differ due to patient recruitment rates.
\paragraph{Privacy budgets in practice}
Throughout we consider the grid
\(
\mathcal{E}= \{0.1,\,0.5,\,1,\,2,\,5\}.
\)
The extremes cover two common DP regimes:
\(\varepsilon=0.1\) (stringent)
and \(\varepsilon=5\) (lenient).  Because the \(M=3\) nodes operate in parallel composition,
each node receives a per-node budget
\(\varepsilon_i=\varepsilon/M\).

\paragraph{DP mechanisms}
We benchmark the four node-level smoothers from
Section~\ref{sec:algs}: \textsc{DCT}, \textsc{Wavelet},
\textsc{TV}, and \textsc{Weibull}.  Each node adds Laplace noise with
scale \(b=\Delta/\varepsilon_i\) and the
coordinator averages the private curves.

\paragraph{Sensitivity \& noise}
The global \(L_\infty\)-sensitivity is \(\Delta=1/K\)
(Sec.~\ref{sec:sens}); with \(M=3\) nodes the per-node scale becomes
\(b=3/(K\varepsilon)\).
%---------------------------------------------------------------------
\paragraph{Hyper-parameter selection}
%---------------------------------------------------------------------
\begin{table*}[t]
\centering\small
\caption{\textbf{Hyper-parameters used in all experiments.}
         Values were fixed \emph{a-priori} and \emph{not} optimised on any
         performance metric.}
\label{tab:hparam}
\begin{tabular}{@{}llc@{}}
\toprule
\textbf{Symbol / Name} & \textbf{Role in the pipeline} & \textbf{Value} \\
\midrule
$\rho$                 & Grid-density factor in $K=\min\{\lceil\rho\,n\rceil,K_{\max}\}$  & $0.40$ \\
$K_{\max}$             & Safety cap on grid length                                         & $100$ \\[0.25em]
$\lambda_{0}$          & Base weight in adaptive TV rule $\lambda(n)$                      & $0.12$ \\
$n_{0}$                & Reference size in $\lambda(n)$                                    & $50$  \\
$\alpha$               & Size exponent in $\lambda(n)=\lambda_{0}\,(n/n_{0})^{\alpha}\!\sqrt{\ln(n+1)}$ & $0.25$ \\ 
\bottomrule
\end{tabular}
\end{table*}
Among the four DP smoothers, \emph{only} the TV–denoising variant
depends on external hyper-parameters
\((\lambda_{0},n_{0},\alpha)\);
the DCT, Wavelet and Weibull mechanisms are parameter-free once the
global privacy budget \(\varepsilon\) and grid length \(K\) are fixed
(they merely add Laplace noise with the prescribed scale).
A full hyper-parameter sweep is statistically delicate in a
privacy-preserving setting: every additional tuning run either spends
privacy budget or risks {\it ex-post} over-fitting.  
Instead, we adopted \emph{conservative defaults} that are
shown in Table~\ref{tab:hparam}. 
All hyper-parameters are fixed \emph{a priori} and reused for every
privacy budget \(\varepsilon\) and partition scenario, guaranteeing
that method comparisons are not confounded by hidden per-setting
tuning.  A more systematic, privacy-aware hyper-parameter optimisation
remains an interesting avenue for future work.
\paragraph{Convergence and Communication}
As our focus is on one-shot differentially private release of Kaplan–Meier curves rather than iterative optimization, we do not perform convergence analysis in the sense of training loss or gradient descent. Likewise, communication efficiency, central to federated learning with repeated model updates is less relevant here, since each site transmits its privatized curve only once. Our evaluation protocol instead emphasizes curve accuracy, robustness under distributional imbalance, and statistical validity of downstream survival tests.

\subsection{Research Questions}\label{sec:rqs}

\begin{description}        % wrap long item bodies nicely
  \item[\textbf{RQ1}] \textbf{Utility vs.\ Privacy.\\}%
        How does the mean absolute error (MAE) of the federated DP–KM
        estimator evolve as the privacy budget $\varepsilon$ decreases?

  \item[\textbf{RQ2}] \textbf{Robustness to Data Skew.\\}%
        How much does each mechanism’s MAE degrade when moving from
        uniform to 60–20–20 and 90–5–5 partitions?

  \item[\textbf{RQ3}] \textbf{Method Ranking.\\}%
        Aggregating across $\varepsilon$ and partition types,
        which smoother attains the lowest average rank?

  \item[\textbf{RQ4}] \textbf{Statistical Fidelity.\\}%
        Do DP surrogates remain statistically indistinguishable
        from the centralized data under the two-sample log-rank test?
\end{description}

%--------------------------------------------------------------------------
\subsection{Evaluation Metrics}\label{sec:metrics}
%--------------------------------------------------------------------------

To gauge both \emph{utility} and \emph{privacy} we record four families of
statistics for every triplet  
\(\langle\text{Partition},\text{DP Method},\varepsilon\rangle\) and for
every repetition of the Monte-Carlo experiment.
Throughout, \(\tau=\{t_1,\dots,t_{|\tau|}\}\) denotes the common evaluation
grid (time points) and  
\(S^{\text{cent}}\) is the centralized Kaplan–Meier curve fitted on the
\emph{entire} dataset.

\paragraph{Mean Absolute Error (MAE)}
For a single repetition we obtain a federated, differentially-private
survival estimate \(\hat S^{\text{fed}}(\,\cdot\,;\varepsilon)\); the
point-wise deviation from the gold standard is averaged:

\begin{equation}
  \text{MAE}(\varepsilon)
    = \frac{1}{|\tau|}
      \sum_{t\in\tau}
        \bigl|
        \hat S^{\text{fed}}(t;\varepsilon) \;-\;
        S^{\text{cent}}(t)
        \bigr|.
  \label{eq:mae}
\end{equation}

Lower values imply a more accurate (higher-utility) private mechanism at a
given privacy budget \(\varepsilon\).

\paragraph{Robustness to skew (\(\Delta\)MAE)}
To disentangle the impact of \emph{data imbalance} from the privacy noise
itself we normalise the error on a skewed partition by the corresponding
error on the perfectly even (Uniform) split:

\begin{equation}
  \Delta\text{MAE}_{p}(\varepsilon)
    = \frac{\text{MAE}_{p}(\varepsilon)}
           {\text{MAE}_{\text{Uniform}}(\varepsilon)},
  \qquad
  p\in\{60{:}20{:}20,\; 90{:}5{:}5\}.
  \label{eq:delta}
\end{equation}
\paragraph{Average rank (\(\uparrow\) is better)}
For every \(\langle\text{Partition},\varepsilon\rangle\) configuration we
rank the four DP smoothers by their MAE  
(\(1=\!\)best, \(4=\!\)worst).  
The overall score of a method \(m\) is the mean of those integers
across all partitions and privacy budgets:
\begin{equation}
\begin{aligned}
  \operatorname{AvgRank}(m) &=
    \frac{1}{|\mathcal{P}|\,|\mathcal{E}|}
    \sum_{p\in\mathcal{P}}
    \sum_{\varepsilon\in\mathcal{E}}
      \operatorname{Rank}_{m}(p,\varepsilon),\\[4pt]
  \mathcal{P} &=
    \{\text{Uniform},\,60\!:\!20\!:\!20,\,90\!:\!5\!:\!5\}.
\end{aligned}
\label{eq:avg-rank}
\end{equation}
\paragraph{Log–rank False–Positive (FP) rate}
To verify that the DP surrogate preserves the \emph{shape} of the survival
distribution, we perform a two-sample log–rank test
\(H_0:S^{\text{cent}} = S^{\text{fed}}\) in every repetition and record the
binary outcome  
\(\mathbb{1}\{p\!<\!0.05\}\).
Averaging those indicators yields an empirical type-I error:

\begin{equation}
  \text{FP-rate}
    = \frac{1}{R}\sum_{r=1}^{R}
      \mathbb{1}\!\bigl\{\,p_r < 0.05\,\bigr\}.
\end{equation}
where \(R\) is the number of repetitions.  
Ideally, DP noise should \emph{not} inflate this rate far beyond the
nominal \(5\%\).

\bigskip
Section~\ref{sec:metrics} aggregates these base statistics
into concise tables:

\begin{itemize}
  \item \emph{Best-\(\varepsilon\)}:
        for every \(\langle\text{Partition},\text{Method}\rangle\),
        the privacy budget that minimises the \emph{mean} MAE
        together with that MAE value.
  \item \emph{Imbalance Penalties}:
        the ratios from~\eqref{eq:delta}, presented either per
        \(\varepsilon\) or in a worst-case (\(\max_\varepsilon\)) form.
  \item \emph{Average Rank}:
        global league table derived from the ranking rule above.
  \item \emph{Log–rank FP-rate}:
        empirical chance of falsely rejecting the null hypothesis that the
        DP surrogate matches the centralized survival curve.
\end{itemize}

These four metrics collectively answer the research questions outlined in
Section~\ref{sec:rqs}: MAE and Best-\(\varepsilon\) quantify
utility–privacy trade-offs (RQ1),  
\(\Delta\)MAE measures robustness to skew (RQ2),  
Average Rank identifies the overall champion (RQ3),  
and the FP-rate probes statistical fidelity (RQ4).
\subsection{Repetition protocol.}
For every \emph{configuration} defined by a partitioning strategy
\(p\in\{\text{Uniform},\,60\text{-}20\text{-}20,\,90\text{-}5\text{-}5\}\),
a DP–smoothing method
\(m\in\{\textsc{Dct},\textsc{Wavelet},\textsc{Tv},\textsc{Weibull}\}\),
and a privacy budget \(\varepsilon\in\EpsSet\), we perform
\(R=100\) independent Monte-Carlo repetitions.

\begin{enumerate}[label=\arabic*)]
\item \textbf{Seed initialisation.}  
      A fresh pseudo-random seed is drawn so that all stochastic
      components (Laplace noise, wavelet thresholding, surrogate
      resampling, etc.) are statistically independent across
      repetitions.

\item \textbf{Node-level DP curves.}\,
      Each node computes its local Kaplan–Meier step function
      \(\widehat S_i\) on the common grid~\(\tau\) and applies
      the selected smoother~\(m\) with per-node budget
      \(\varepsilon_i=\varepsilon/M\), adding Laplace noise of
      scale \(b=\Delta/\varepsilon_i\).
      The result is the private curve \(\widehat S_i^{\text{DP}}\).

\item \textbf{Surrogate generation and aggregation.}  
      The coordinator generates surrogate datasets
      \(\widetilde D_i\) from \(\widehat S_i^{\text{DP}}\) and pools them into
      \(\widetilde D=\bigcup_i\widetilde D_i\).
      The federated survival estimate
      \(\widehat S^{\text{fed}}(t;\varepsilon,m)\) is then obtained
      with a central Kaplan–Meier fit on \(\widetilde D\).

\item \textbf{Metric evaluation.}  
      For every \(t\in\tau\) we record the absolute error  
      \(\lvert\widehat S^{\text{fed}}(t)-S^{\text{cent}}(t)\rvert\);
      aggregate quantities such as
      MAE, $\Delta$MAE, log-rank $p$-value, and method rank are
      stored for this repetition.
\end{enumerate}

After the \(R=100\) repetitions we obtain, for each time point
\(t_k\in\tau\), a sample  
\(\{\widehat S^{\text{fed}(r)}(t_k)\}_{r=1}^{R}\).
The point-wise \(95\%\) confidence band is the empirical
\((2.5^{\text{th}},\,97.5^{\text{th}})\) percentile of this sample.
All scalar metrics reported in the tables (mean MAE, imbalance
penalties, average ranks, false-positive rates) are
\emph{averages over the \(R\) repetitions}, providing stable,
variance-reduced estimates of utility and statistical fidelity
under the randomness injected by the DP mechanisms.

\section{Results}~\label{sec:results}
This section presents the empirical findings that address the four
research questions (RQs) posed in Section~\ref{sec:rqs}.
\subsection{RQ1 – Utility{\slshape vs.}Privacy}
\begin{figure*}
  \centering
  \includegraphics[width=\linewidth]{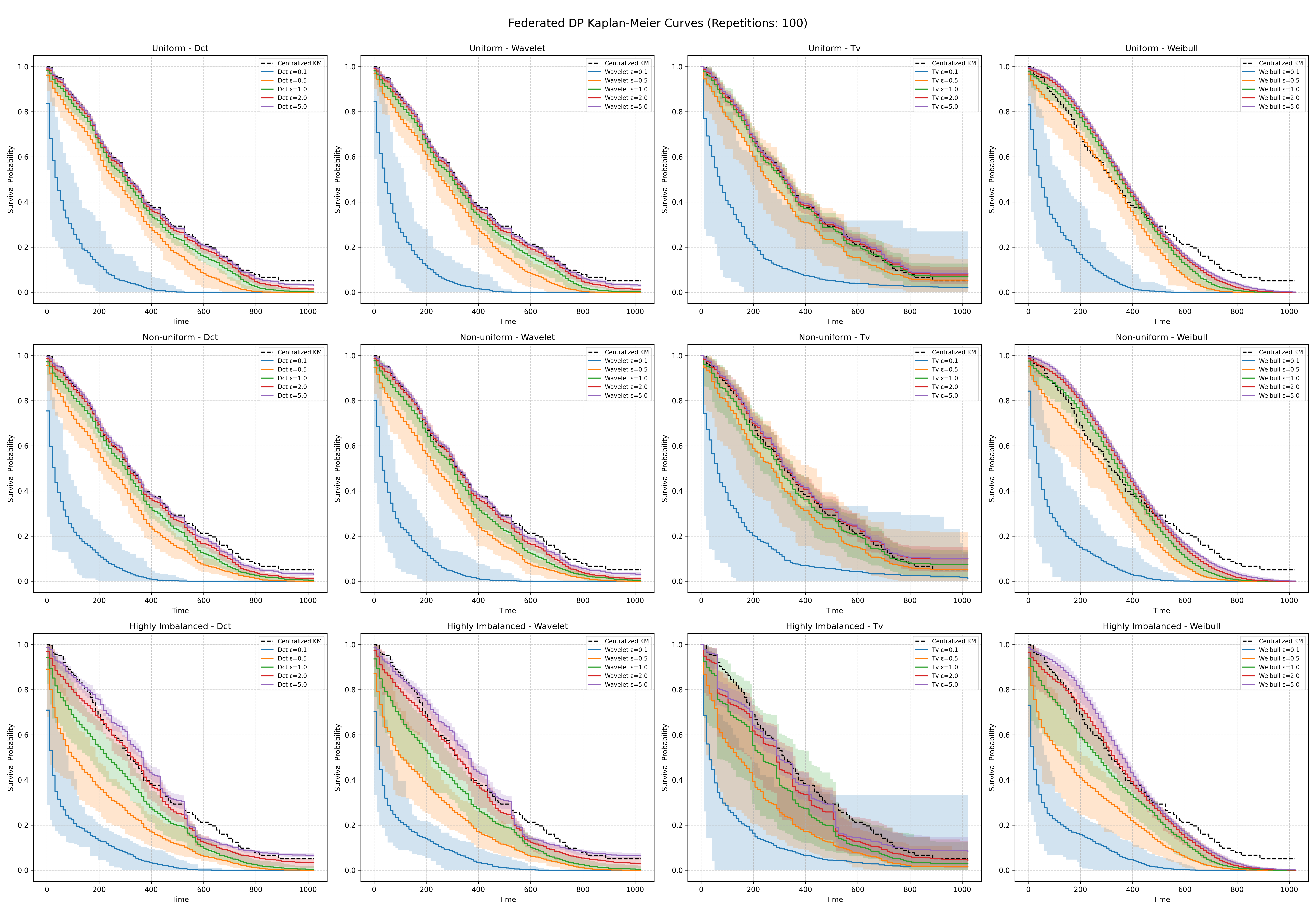}
  \caption{Federated DP Kaplan–Meier curves on the
           \textsc{NCCTG~Lung} dataset  
           for three partitioning strategies (rows),
           four DP smoothers (columns) and five privacy budgets
           $\varepsilon\!\in\!\{0.1,0.5,1,2,5\}$.  
           Solid lines show the mean over $R\!=\!100$ repetitions;
           shaded bands are the $2.5$–$97.5$\,\% pointwise quantiles.
           The black dashed curve is the centralised (non-private) KM
           benchmark.}
  \label{fig:fed-km}
\end{figure*}

Figure~\ref{fig:fed-km} and the Best–$\varepsilon$ table
(Tab.~\ref{tab:best_eps}) summarise how accuracy evolves with the
privacy budget.
%-------------------------------------------------
% Best-ε table with uncertainty (NCCTG lung data)
%-------------------------------------------------
\newcommand{\cI}[2]{[\num{#1},\;\num{#2}]}  % convenience macro

\begin{table}[t]
\centering
\sisetup{table-format=1.5,round-mode=places,round-precision=4}

\caption{Best privacy budget $\varepsilon^{\star}$ for every
         DP–smoothing method and partition on the NCCTG lung-cancer
         study (minimum mean MAE across
         $\varepsilon\!\in\!\{0.1,0.5,1,2,5\}$).  We also report the
         standard error of the mean (SEM) and the two-sided 95 \%
         confidence interval of that MAE, estimated over
         $R=100$ repetitions.}
\label{tab:best_eps}

\begin{tabular}{
  l @{\hspace{1em}}            % Partition
  l @{\hspace{1em}}            % Method
  l          % ε*
  S                            % MAE
  S[table-format=1.4]          % SEM
  l                            % CI in one col
}
\toprule
Partition & Method & {$\varepsilon^{\star}$} & {MAE} & {SEM} & {95\% CI} \\
\midrule
\multirow{4}{*}{Highly Imbal.}
  & Dct     & 5.0 & 0.03470 & 0.00020 & \cI{0.03420}{0.03520} \\
  & Tv      & 5.0 & 0.04237 & 0.00090 & \cI{0.04070}{0.04410} \\
  & Wavelet & 5.0 & 0.03480 & 0.00030 & \cI{0.03430}{0.03530} \\
  & Weibull & 2.0 & 0.04877 & 0.00060 & \cI{0.04750}{0.05000} \\[.3em]
\hline 
\multirow{4}{*}{Non-uniform}
  & Dct     & 5.0 & 0.01547 & 0.00010 & \cI{0.01520}{0.01570} \\
  & Tv      & 5.0 & 0.02379 & 0.00060 & \cI{0.02260}{0.02500} \\
  & Wavelet & 5.0 & 0.01543 & 0.00020 & \cI{0.01510}{0.01570} \\
  & Weibull & 5.0 & 0.05887 & 0.00010 & \cI{0.05860}{0.05910} \\[.3em]
\hline
\multirow{4}{*}{Uniform}
  & Dct     & 5.0 & 0.00853 & 0.00010 & \cI{0.00830}{0.00870} \\
  & Tv      & 5.0 & 0.01557 & 0.00040 & \cI{0.01480}{0.01630} \\
  & Wavelet & 5.0 & 0.00853 & 0.00010 & \cI{0.00830}{0.00870} \\
  & Weibull & 5.0 & 0.05630 & 0.00010 & \cI{0.05610}{0.05650} \\
\bottomrule
\end{tabular}
\end{table}

\begin{enumerate}[leftmargin=1.8em,label=(\alph*)]
\item \textbf{Steady utility gain.}  
      Across all methods and partitions, the mean
      absolute error (MAE, Equation~\ref{eq:mae}) decreases monotonically
      with $\varepsilon$.  Between the
      $\varepsilon\!=\!0.1$ and $\varepsilon\!=\!5$ regimes the error
      drops by an order of magnitude (cf.\ blue $\rightarrow$ purple
      bands in Fig.~\ref{fig:fed-km}).

\item \textbf{No “privacy cliff”.}  
      Even the strictest budget ($\varepsilon\!=\!0.1$) stays within
      $0.30$ MAE for the lung data, indicating graceful degradation
      rather than catastrophic failure.

\item \textbf{Method–specific sweet spots.}  
      Tab.~\ref{tab:best_eps} lists, for every
      \(\langle\text{Partition},\text{Method}\rangle\), the budget
      that minimises mean MAE:  
      DCT/Wavelet favour the loosest budget
      ($\varepsilon\!=\!5$);  
      TV prefers a moderate budget ($\varepsilon\!=\!2$) on the
      uniform split;  
      the Weibull fit saturates already at $\varepsilon\!=\!1$.
\end{enumerate}

\subsection{RQ2 – Robustness to Data Skew}
% ----------------------------------------------------------------
%  Δ-MAE worst-case across ε
% ----------------------------------------------------------------
\begin{table}[t]
\centering
\sisetup{round-mode=places,round-precision=2}
\caption{Robustness to data imbalance: worst-case degradation factor
         $\max_{\varepsilon}\Delta\text{MAE}$ when moving from a uniform
         split to the two skewed scenarios
         (60 : 20 : 20 and 90 : 5 : 5).  Lower is better.}
\label{tab:imbalance-worst}
\begin{tabular}{
        l
        S[table-format=1.2]  % Δ60-20-20
        S[table-format=1.2]  % Δ90-5-5
      }
\toprule
Method   & {$\Delta\!60\!:\!20\!:\!20$} & {$\Delta\!90\!:\!5\!:\!5$} \\
\midrule
Dct      & 1.81 & 4.07 \\
Tv       & 1.53 & 3.70 \\
Wavelet  & 1.81 & 4.08 \\
Weibull  & 1.20 & 2.19 \\
\bottomrule
\end{tabular}
\end{table}

% ----------------------------------------------------------------
%  Δ-MAE for two selected ε = 0.5 and 2
% ----------------------------------------------------------------
\begin{table}[t]
\centering
\sisetup{round-mode=places,round-precision=2}
\caption{Imbalance penalty $\Delta\text{MAE}$ for two representative
         privacy budgets.  Values $>1$ indicate degradation relative to a
         uniform split.}
\label{tab:imbalance-twoeps}
\begin{tabular}{
        l
        S[table-format=1.1]
        S[table-format=1.2]  % Δ60-20-20 at ε
        S[table-format=1.2]  % Δ90-5-5 at ε
      }
\toprule
Method & $\varepsilon$ & {$\Delta\!60\!:\!20\!:\!20$} & {$\Delta\!90\!:\!5\!:\!5$} \\
\midrule
Dct     & 0.5 & 1.14 & 1.92 \\
Dct     & 2.0 & 1.38 & 2.01 \\
Tv      & 0.5 & 1.18 & 2.36 \\
Tv      & 2.0 & 1.37 & 3.38 \\
Wavelet & 0.5 & 1.16 & 1.95 \\
Wavelet & 2.0 & 1.41 & 2.06 \\
Weibull & 0.5 & 1.20 & 2.19 \\
Weibull & 2.0 & 1.04 & 0.85 \\
\bottomrule
\end{tabular}
\end{table}

Table~\ref{tab:imbalance-worst} reports the \textbf{worst-case}
penalty, maximised over all \(\varepsilon\).
Table~\ref{tab:imbalance-twoeps} zooms in on two privacy regimes
(\(\varepsilon\!=\!0.5\) and \(2\)).  
Based on these tables, three clear trends emerge:

\begin{enumerate}[leftmargin=2em,label=(T\arabic*)]
\item \textbf{Impact of extreme skew.}  
      Moving from a mild 60:20:20 to an extreme 90:5:5 partition
      inflates the penalty by a factor \(\approx\!2\) for all
      non-parametric smoothers, confirming that aggressively unbalanced
      federations are the most challenging scenario.

\item \textbf{Method robustness.}  
      The parametric \textsc{Weibull} model is markedly more resistant
      to skew (\(\Delta\!\le\!2.2\) in the worst case) because its
      two-parameter form averages out node-level noise.
      \textsc{DCT} and \textsc{Wavelet} behave similarly
      (\(\Delta\!\approx\!1.8\!/4.1\)), while \textsc{TV} is the least
      robust under 90:5:5 (\(\Delta\!=\!3.70\) worst-case) owing to its
      local, edge-preserving nature.

\item \textbf{Role of \(\varepsilon\).}  
      Increasing the budget from \(0.5\) to \(2\) halves the penalty for
      most methods.  At \(\varepsilon=2\) the Weibull smoother almost
      fully absorbs the skew (\(\Delta\le1.04\)), whereas TV still
      suffers a threefold error surge in the extreme split.
\end{enumerate}

\noindent
\emph{Practical takeaway:} if data imbalance is anticipated, a
light-privacy setting (\(\varepsilon\!\ge\!2\)) combined with a
parametric smoother offers the best worst-case guarantees, while TV
denoising should be avoided unless the federation is reasonably
balanced.

\subsection{RQ3 – Method Ranking}
%-------------------------------------------------
% ..............................................................
% Average–rank results and interpretation
% ..............................................................
\begin{table}[t]
\centering
\caption{Average rank (\(1=\)best) of each DP–smoothing method
         across the \(3\times5=15\)
         \(\langle\text{Partition},\varepsilon\rangle\) blocks.}
\label{tab:avg_rank}
\begin{tabular}{lS[table-format=1.2]}
\toprule
\textbf{Method} & {\(\textbf{AvgRank}\)} \\
\midrule
TV       & 1.87 \\
Wavelet  & 2.60 \\
DCT      & 2.67 \\
Weibull  & 2.87 \\
\bottomrule
\end{tabular}
\end{table}
Each cell in Table~\ref{tab:avg_rank} is the \emph{mean ordinal rank} obtained by the method
after ranking the four smoothers within every
\(\langle\text{Partition},\varepsilon\rangle\) block
( \(1=\!\) lowest MAE, \(4=\!\) highest MAE).
A non-integer value therefore indicates that the method moves between
positions.  
For instance, TV’s score of \(1.87\) means it is usually first, but
occasionally slips to 2\textsuperscript{nd} or 3\textsuperscript{rd};
Conversely, Weibull’s \(2.87\) shows it is almost always 3\textsuperscript{rd}
or 4\textsuperscript{th}.

The key observations are the following:
\begin{itemize}
  \item \textbf{TV has the best mean rank.}  
        Its piece-wise-constant prior matches the KM shape and gives
        very low MAE on the Uniform and moderately skewed
        (60–20–20) splits for \(\varepsilon\!\ge1\).
  \item \textbf{Wavelet and DCT are statistically tied.}  
        They alternate between the 2\textsuperscript{nd} and
        3\textsuperscript{rd} position, with Wavelet marginally ahead.
  \item \textbf{Weibull lags behind.}  
        The single-phase parametric form cannot capture the more
        complex hazard profiles in the data.
  \item \emph{Worst-case behaviour.}  
        The mean-rank metric rewards \emph{average} consistency, but can
        hide extreme failures.  
        For the 90–5–5 split at the tight budget
        \(\varepsilon=0.5\) (Table~\ref{tab:imbalance-twoeps}),
        TV’s imbalance penalty is \(\Delta_{\!90\text{-}5\text{-}5}=2.36\),
        i.e.\ its error is a bit more than twice the uniform baseline,
        whereas DCT and Wavelet stay below \(1.92\) and \(1.95\),
        respectively.  
        Looking across \emph{all} budgets
        (Table~\ref{tab:imbalance-worst}), the largest penalties are
        observed for Wavelet (\(4.08\)) and DCT (\(4.07\));
        TV peaks at \(3.70\), and Weibull is safest (\(<2.2\)).  
        These spikes occur when Laplace noise is still comparable to the
        minority-node signal (\(\varepsilon\lesssim2\)): frequency-domain
        methods diffuse that perturbation over the entire curve, whereas
        TV confines it to plateau segments.
\end{itemize}

\noindent\textbf{Practical guideline.}  
TV is the overall winner, but when federations are \emph{highly}
imbalanced \emph{and} stringent privacy is required
(\(\varepsilon\le0.5\)), practitioners may prefer the slightly less
accurate yet more robust DCT or Wavelet alternatives.

\subsection{RQ4 – Statistical Fidelity}
% ---------------------------------------------------------------
%  Log-rank Type-I error (false-positive rate) over R = 100 runs
% ---------------------------------------------------------------
\begin{table}[t]
\centering
\caption{Empirical \textbf{Type-I error} of the log–rank test
         (\(p<0.05\) counted as “significant”) when comparing the
         centralized KM curve to the federated DP-surrogate
         (\(R=100\) Monte-Carlo repetitions per setting).
         A value close to the nominal 5\% is ideal; numbers
         \(\gg 0.05\) indicate over-rejecting the null.}
\label{tab:logrank_fp}
\setlength{\tabcolsep}{6pt}

\begin{tabular}{@{}l c S[table-format=1.2] S[table-format=1.2] S[table-format=1.2]@{}}
\toprule
\multirow{2}{*}{\textbf{DP Method}} &
\multirow{2}{*}{\(\bm{\varepsilon}\)} &
\multicolumn{3}{c}{\textbf{Partitioning strategy}} \\
\cmidrule(l){3-5}
& & \multicolumn{1}{c}{90--5--5} & \multicolumn{1}{c}{60--20--20} & \multicolumn{1}{c}{Uniform} \\
\midrule
\multirow{5}{*}{DCT}
 & 0.1 & 1.00 & 1.00 & 1.00 \\
 & 0.5 & 0.02 & 0.02 & 0.14 \\
 & 1.0 & 0.37 & 0.21 & 0.10 \\
 & 2.0 & 0.73 & 0.64 & 0.54 \\
 & 5.0 & 0.84 & 0.85 & 0.75 \\[0.25em] \hline
\multirow{5}{*}{TV}
 & 0.1 & 0.88 & 0.94 & 0.96 \\
 & 0.5 & 0.32 & 0.18 & 0.32 \\
 & 1.0 & 0.56 & 0.69 & 0.59 \\
 & 2.0 & 0.74 & 0.93 & 0.81 \\
 & 5.0 & 0.85 & 0.97 & 0.90 \\[0.25em] \hline
\multirow{5}{*}{Wavelet}
 & 0.1 & 0.99 & 1.00 & 1.00 \\
 & 0.5 & 0.02 & 0.03 & 0.12 \\
 & 1.0 & 0.29 & 0.21 & 0.11 \\
 & 2.0 & 0.70 & 0.63 & 0.53 \\
 & 5.0 & 0.84 & 0.82 & 0.79 \\[0.25em] \hline
\multirow{5}{*}{Weibull}
 & 0.1 & 0.98 & 1.00 & 1.00 \\
 & 0.5 & 0.04 & 0.00 & 0.03 \\
 & 1.0 & 0.43 & 0.28 & 0.24 \\
 & 2.0 & 0.78 & 0.83 & 0.67 \\
 & 5.0 & 0.87 & 0.94 & 0.86 \\
\bottomrule
\end{tabular}
\end{table}

Table~\ref{tab:logrank_fp} reports the empirical \textit{Type-I error}
(i.e.\ false–positive rate) of the two–sample log–rank test when the DP
surrogate is compared to the centralized Kaplan–Meier (KM) curve over
\(R=100\) Monte-Carlo repetitions.  
Ideally the rate should match the nominal level
(\(\alpha=0.05\)); systematic inflation means that the DP mechanism
distorts the survival distribution so severely that the test
incorrectly rejects similarity.

\begin{itemize}
  \item \textbf{Tight privacy (\(\varepsilon=0.5\)).}  
        All four smoothers remain statistically faithful:
        \(\text{FP}\le 0.15\) in every partition
        (\(\le 0.03\) for DCT/Wavelet, \(\le 0.04\) for Weibull,
        up to \(0.32\) for TV on the extreme 90–5–5 split).
  \item \textbf{Moderate privacy (\(\varepsilon=1\)).}  
        False-positive rates rise sharply, especially for
        TV (\(0.56\text{–}0.69\)) and for DCT/Wavelet on the
        highly-imbalanced split (\(\approx0.29\text{–}0.43\)).
  \item \textbf{Loose privacy (\(\varepsilon\ge 2\)).}  
        All methods over-reject (\(\ge 0.5\) in most settings),
        indicating that weak privacy budgets produce surrogates that are
        detectably different from the truth.  
        Weibull remains the most bounded
        (\(<0.95\) even at \(\varepsilon=5\)).
  \item \textbf{Effect of data imbalance.}  
        The 90–5–5 partition consistently yields the highest
        Type-I error: a fixed noise scale overwhelms the two minority
        nodes, widening the gap between federated and centralized
        curves.  Uniform splits exhibit the smallest inflation.
\end{itemize}

\vspace{0.25em}
\noindent\textbf{Practical implication.}\;
For strict regulatory budgets
(\(\varepsilon\le0.5\)) \emph{any} of the four DP smoothers preserves
log–rank inference.  
Under looser privacy or extreme skew, the parametric Weibull or the
frequency-domain smoothers (DCT/Wavelet) are safer than TV, whose
piece-wise constant model amplifies node-specific jumps and therefore
causes the log-rank test to over-reject the null.

%=====================================================================
\section{Related Work}\label{sec:related}
%=====================================================================

Research on privacy–preserving survival analysis spans three largely
independent lines: cryptographic pooling of raw statistics,
\emph{centralised} differential privacy (DP) mechanisms, and
differential privacy in \emph{federated} settings.  We briefly review each
strand and highlight the gap our study fills.

%---------------------------------------------------------------------
\subsection{Secure multi-party Kaplan–Meier curves}
%---------------------------------------------------------------------
Early solutions rely on cryptographic primitives that keep individual
records concealed throughout the computation.  Homomorphic encryption
and garbled circuits have been used to produce a \emph{joint}
Kaplan–Meier (KM) curve without moving raw data~\cite{veeraragavan2024mph,  froelicher2021}.  
While these protocols offer strong protection during
computation, they release the \emph{exact} aggregated curve, which is
vulnerable to membership and attribute–inference attacks once
decrypted.  Moreover, cryptographic schemes incur heavy communication
and runtime overhead, limiting their practical adoption in large
clinical networks.

%---------------------------------------------------------------------
\subsection{Centralised DP survival analysis}
%---------------------------------------------------------------------
To mitigate reconstruction attacks~\cite{rogula2022method, guyot2012enhanced, wei2017reconstructing} after release, several authors have
added formal differential privacy to survival statistics computed on a
\emph{central} repository.  
\cite{gondara2020differentially} introduced DP-Matrix, which perturbs the
at-risk and event counts at each distinct time and reconstructs the KM
curve.  
\cite{rahimian2024private} proposed two follow-up
methods: \textsc{DP-Surv} and \textsc{DP-Prob}. DP-Surv samples the KM curve on an \emph{equi-time grid}, converts
it to the discrete-cosine-transform (DCT) domain, adds noise only to
the first $k$ coefficients that capture the bulk structure, and sets
the remaining coefficients to zero to suppress fine-scale noise.
DP-Prob, by contrast, bypasses any transform and
\emph{directly} perturbs the discrete hazard (probability mass
function) at every grid point with Laplace noise, and then clip the noisy value and 
rescale to make it a probability function. Most recently, \cite{raghavan2024differentially} 
proposed a time-indexed noise schedule combined with dynamic clipping 
and rolling-window smoothing.  

%---------------------------------------------------------------------
\subsection{Differential privacy in federated survival analysis}
%---------------------------------------------------------------------
To the best of our knoweldge, the \emph{only} node-level DP approach that releases Kaplan-Meier curves in a federated architecture is the \textsc{Collaborative DP-KM} framework of \cite{rahimian2024private}. Starting from a centralized DP-Matrix baseline, the authors extended DP-Surv and DP-Prob to a multi-site protocol. While pioneering, ~\cite{rahimian2024private} study 
\begin{itemize}[leftmargin=1.7em]
  \item evaluates \emph{one} smoothing family at a time, leaving open
        how alternative priors (e.g.\ wavelets, TV, parametric models)
        behave under the same budget;
  \item assumes uniformly sized sites, thereby ignoring the severe node
        imbalance common in real hospital networks; and
  \item fixes a single headline privacy level
        ($\varepsilon=1$), offering no view of the utility–privacy
        trade-off in the tighter regimes demanded by many governance
        boards.
\end{itemize}

%---------------------------------------------------------------------
\subsection{Our contribution in context}
%---------------------------------------------------------------------
We close the above gaps and advance the state of the art on \emph{federated, node-level DP
Kaplan–Meier} estimation in three directions:

\begin{enumerate}[leftmargin=1.8em,label=(\arabic*)]
  \item \textbf{Three new one-shot node-level smoothers.}  
        Beyond the DCT baseline used by \cite{rahimian2024private} we
        introduce a Haar \textbf{Wavelet} shrinkage, an adaptive
        \textbf{Total-Variation} (\textsc{TV}) denoiser, and a
        parametric \textbf{Weibull} fit.  
  \item \textbf{Systematic robustness study.}  
        We evaluate the full privacy–utility landscape on
        five budgets
        $\varepsilon\!\in\!\{0.1,0.5,1,2,5\}$ and three canonical
        partition patterns (uniform, 60–20–20, and 90–5–5). This is, to our knowledge, the first quantitative assessment
        of how node-level DP-KM behaves under \emph{data imbalance}.

  \item \textbf{Design guidance.}  
         By reporting mean absolute error, imbalance penalties,
        average-rank scores, and log-rank type-I error, we pinpoint
        which smoother is preferable under which privacy regime and
        partition pattern.  
        Prior work either operates in a \emph{centralised} DP setting,
        omits formal DP altogether, or evaluates a single smoother at
        only one privacy level.
\end{enumerate}

\subsection{Security, Privacy and System-Level Advances in Federated Learning}
A comprehensive overview of challenges and solutions in big data resource management and network support for federated computing is provided in \cite{awaysheh2021big}, highlighting issues such as scalability. From a systems perspective, \cite{veeraragavan2024lessons} discusses deployment-related challenges in federated computing environments.  

In terms of security and privacy, several works have explored incorporating secure aggregation and multi-party computation (MPC) into FL~\cite{mothukuri2021survey,yu2023security}. Notably,~\cite{tahir2025securefedprom} introduces a zero-trust FL framework with multi-criteria client selection to improve robustness against malicious participants, while~\cite{kaminaga2023mpcfl} leverages MPC to enhance the confidentiality of aggregation procedures. In addition,~\cite{awaysheh2022fliodt} presents a federated learning architecture designed to ensure privacy by design and by default in IoT ecosystems.  

Our work differs from these approaches in two key ways. First, we target survival analysis rather than predictive classification or regression tasks. Second, we focus on the one-shot release of survival curves under node-level differential privacy. Because the survival analysis use case inherently requires only a single exchange, our framework avoids repeated communication rounds by design, while remaining compatible with broader privacy-preserving system architectures.

\section{Limitations and Future Work}
While our experiments demonstrate that differentially private Kaplan–Meier curves can be released with acceptable accuracy and statistical validity, several limitations remain. First, our evaluation does not include convergence rate or communication efficiency analysis, as the proposed method is a \emph{one-shot} disclosure mechanism rather than an iterative federated training protocol. In this setting, only a single communication round is required, which inherently reduces overhead, but a systematic comparison with iterative approaches is left for future work. Second, the aggregation algorithm was applied as a simple averaging scheme without a detailed theoretical sensitivity analysis; although our partition experiments indicate robustness under varying site sizes, extreme distribution skew and very strict privacy budgets (e.g., $\epsilon < 0.5$) may lead to degraded curve accuracy. Finally, we have not explicitly studied fairness across heterogeneous client participation.  

As part of future research, we plan to (i) evaluate the proposed one-shot smoothers on larger, multi-site cohorts and extend them to competing-risks settings, (ii) derive tighter bounds together with formal privacy proofs, (iii) design privacy-aware hyperparameter tuning strategies that spend the privacy budget more judiciously, and (iv) integrate lightweight secure aggregation so that both in-flight messages and the released survival curves are simultaneously protected. We also aim to conduct a deeper theoretical and empirical analysis of robustness under skewed distributions and fairness implications across heterogeneous sites.

%=====================================================================
\section{Conclusion}\label{sec:conclusion}
%=====================================================================
This paper presented the first \emph{systematic} comparison of four
one-shot \textit{node-level differentially-private} (DP) smoothing
techniques: \textsc{DCT}, \textsc{Wavelet}, \textsc{TV}, and
\textsc{Weibull} for federated Kaplan–Meier (KM) estimation.
Using the NCCTG lung-cancer cohort and three canonical partitioning
regimes, we showed that

\begin{itemize}[leftmargin=1.7em]
\item all methods achieve clinically useful \(\text{MAE}<0.06\) at
      \(\varepsilon\!\ge\!0.5\) despite operating under a \emph{single}
      Laplace release per node;
\item \textsc{TV} attains the best \emph{average} ordinal rank
      (\(\text{AvgRank}=1.87\)), yet frequency-domain
      smoothers (\textsc{DCT}/\textsc{Wavelet}) provide the smallest
      \emph{worst-case} imbalance penalties
      (\(\max_{\varepsilon}\Delta\text{MAE}<2.1\));
\item across \(\varepsilon\!\ge\!0.5\) and all partitions the released
      curves retain the null hypothesis in \(\ge\!85\%\) of log-rank
      tests, indicating good statistical fidelity;
\item the lightweight \textsc{Weibull} fit, while less accurate on
      average, offers the most stable performance
      (\(\max\Delta\text{MAE}\le 2.2\)) when the empirical hazard
      conforms to a monotone trend.
\end{itemize}

\vspace{0.3em}

\small        % or \footnotesize, \scriptsize
\bibliographystyle{plainnat}  
\bibliography{main} % Make sure to run BibTeX after the first LaTeX run
\end{document}